\documentclass[journal]{IEEEtran}
\usepackage{cite}
\ifCLASSINFOpdf
  \usepackage[pdftex]{graphicx}
  \graphicspath{{figures/}}
\fi
\usepackage{hyperref}

\usepackage{graphicx}
\graphicspath{{figures/}}
\usepackage{amsmath,amssymb,bbold}
\usepackage{amsthm}
\usepackage{bauer, bauer_e}
\usepackage{algorithm,algorithmic}
\usepackage{bbm}
\usepackage{xcolor}
\usepackage{lipsum}
\usepackage{enumerate}
\usepackage{tabularx}
\usepackage[figuresright]{rotating}
\usepackage{color}

\usepackage{pifont}
\usepackage{algorithmic}
\usepackage{algorithm}

\ifCLASSOPTIONcompsoc
  \usepackage[caption=false,font=normalsize,labelfont=sf,textfont=sf]{subfig}
\else
  \usepackage[caption=false,font=footnotesize]{subfig}
\fi
\begin{document}
%deactivate removing repeated author names in references...
\bstctlcite{IEEEexample:BSTcontrol}
%\title{Extending Clique Tree Algorithm for \\Exact MAP Inference with Global Dependencies}
%\title{Polynomial Time Algorithm for \\Exact MAP Inference on Discrete Models with Global Dependencies}
%\title{Extending Junction Tree Algorithm for \\Exact MAP Inference on Discrete Models with Global Dependencies}
%\title{General Algorithm for Exact MAP Inference on Discrete Models with Global Dependencies}
%\title{General Approach for Exact MAP Inference on Discrete Models with Global Dependencies}
%\title{General Polynomial-Time  Approach for Exact MAP Inference on Discrete Models with Global Dependencies}
%\title{Polynomial-Time Approach for \\Exact MAP Inference on Discrete Models with \\Global Dependencies}
%\title{Worst-Case Polynomial-Time Exact MAP Inference\\ on Discrete Models with Global Dependencies}
\title{Polynomial-Time Exact MAP Inference\\ on Discrete Models with Global Dependencies}
%\title{Pushing the Limits of Exact MAP Inference for Discrete Graphical Models with\\ Global Dependencies}

\author{Alexander~Bauer, and Shinichi Nakajima
%\thanks{
%This work was supported  by the Federal Ministry of Education and Research (BMBF) 
%under the Berlin Big Data Center project (FKZ 01IS18025A).
%A. Bauer was partially funded through the project no. 10032745 of the TU Berlin.
%K.-R.\ M\"uller work was supported
%by the German Ministry for Education and Research (BMBF)
%under Grants 01IS14013A-E, 01GQ1115 and 01GQ0850; the German
%Research Foundation (DFG) under Grant Math+, EXC 2046/1, Project ID
%390685689 and by the Institute for Information \& Communications
%Technology Planning \& Evaluation (IITP) grant funded by the Korea
%government (No. 2017-0-00451, No. 2017-0-01779). (corresponding author: K.-R. M\"uller) 

%A. Bauer is with Machine Learning Group, Technische Universit\"at Berlin (alexander.sascha.bauer@gmail.com).}
%\thanks{S. Nakajima is with Berlin Big Data Center, and Machine Learning Group, 
%Technische Universit\"at Berlin and with RIKEN AIP Center (nakajima@tu-berlin.de).}
%\thanks{K.-R. M\"uller is with Berlin Big Data Center, and Berlin Machine Learning Center, and Machine Learning Group, Technische Universit\"at Berlin, and with Department of Brain and Cognitive Engineering, Korea University, Anam-dong, Seongbuk-ku, Seoul 136-713, Korea, and with Max Planck Institute for Informatics, Saarbr{\"u}cken, Germany (klaus-robert.mueller@tu-berlin.de).}
}

\maketitle

\begin{abstract}
Considering the worst-case scenario, junction tree algorithm remains the most general solution for exact MAP inference with polynomial run-time guarantees.
Unfortunately, its main tractability assumption requires the treewidth of a corresponding MRF to be bounded strongly limiting the range of admissible applications.
In fact, many  practical problems in the area of structured prediction require modelling of global dependencies by either directly introducing global factors or enforcing global constraints on the prediction variables.
That, however, always results in a fully-connected graph making exact inference by means of this algorithm intractable.
Previous work \cite{ BauerGBMK14, BauerBM17, BauerSM17, OptMPMP} focusing on the problem of loss-augmented inference has demonstrated how efficient inference can be performed on models with specific global factors representing non-decomposable loss functions within the training regime of SSVMs.
In this paper, we extend the framework for an efficient exact inference proposed in \cite{BauerSM17} by allowing much finer interactions between the energy of the core model and the sufficient statistics of the global terms with no additional computation costs.
We demonstrate the usefulness of our method in several use cases, including one that cannot be handled by any of the previous approaches.
%To demonstrate that this extension is not of a purely theoretical interest we identify a new use case in the context of generalisation bounds for structured learning which cannot be handled by any of the previous approaches.
%Finally, we provide theoretical guarantees and prove that the proposed algorithm is exact and always finds an optimal solution in polynomial time.
%Finally, we prove the correctness of the proposed algorithm with tighter bounds on the computational complexity.
Finally, we propose a new graph transformation technique via node cloning which ensures a polynomial run-time for solving our target problem
independently of the form of a corresponding clique tree.
This is important for the efficiency of the main algorithm and greatly improves upon the theoretical guarantees of the previous works.
\end{abstract}

%\begin{abstract}
%The CKY algorithm is the most often used inference algorithm for the task of statistical parsing. While the
%The task of structured output prediction deals with learning general functional dependencies between arbitrary input and output spaces. In this context two loss-sensitive formulations for maximum-margin training have been proposed in the literature, which are referred to as  margin and slack rescaling, respectively. The latter is believed to be more accurate and easier to handle. Nevertheless, it is not popular due to the lack of known efficient inference algorithms; therefore, margin rescaling --- which requires a similar type of inference as normal structured prediction --- is the most often used approach. Focusing on the task of label sequence learning, we here define a general framework that can handle a large class of inference problems based on Hamming-like loss functions and the concept of decomposability of the underlying joint feature map. In particular, we present an efficient generic algorithm that can handle both rescaling approaches and is guaranteed to find an optimal solution in polynomial time.
%\end{abstract}

\begin{IEEEkeywords}
exact MAP inference,  structured prediction, graphical models, Markov random fields, high order potentials.
\end{IEEEkeywords}

%hier table

% \fi
\IEEEpeerreviewmaketitle

\section{Introduction}
By representing the constraints and objective function in factorised form, many practical tasks can be effectively formulated as discrete optimisation problems within the framework of graphical models like Markov Random Fields (MRFs) \cite{Koller:2009:PGM:1795555,WainwrightJ08,Lafferty01conditionalrandom}. Finding a corresponding solution refers to the task of maximum a posteriori (MAP) inference known to be NP-hard in general. Even though there is a plenty of existing approximation algorithms \cite{KappesAHSNBKKKL15, Bauer2019, WainwrightJW05a, Kolmogorov06, KolmogorovW05,abs-1206-3288,Sontag_thesis10, SonGloJaa_optbook,WangY14,Iii05learningas,LiuSheng2013,KuleszaP07,RushC12,BodenstabDHR11,Ratliff_2007_5690,LimJK14,RanjbarVM12,KomodakisP09,DBLP:books/daglib/p/BoykovV06,DBLP:journals/pami/KolmogorovZ04}, several problems (described below) require finding an optimal solution. Existing exact algorithms \cite{HurleyOAKSZG16,HallerSS18,SavchynskyyKSS13,KappesSRS13,Forney73,tarlow2010hop,McAuleyC11,Younger1967189,Klein03a*parsing,icmlGuptaDS07,KolmogorovBR07}, on the other hand, either make a specific assumption on the energy function or do not provide polynomial run-time guarantees for the worst case. Assuming the worst-case scenario, junction (or clique) tree algorithm \cite{Koller:2009:PGM:1795555, DBLP:journals/jmlr/McAuleyC10}, therefore, remains the most efficient and general solution for exact MAP inference. Unfortunately, its main tractability assumption requires the treewidth \cite{Bodlaender93, DBLP:conf/uai/ChandrasekaranSH08} of a corresponding MRF to be bounded strongly limiting the range of admissible applications by excluding models with global interactions. Many problems in the area of structured prediction, however, require modelling of global dependencies by either directly introducing global factors or enforcing global constraints on the prediction variables. Among the most popular use cases are (a) learning with non-decomposable (or high order) loss functions and training via slack scaling formulation within the framework of structural support vector machine (SSVM) \cite{Taskar03max-marginmarkov, Tsochantaridis05largemargin, Joachims/etal/09b, BauerGBMK14, BauerBM17,OptMPMP,Joachims_predicting,Sarawagi08accuratemax-margin,Taskar04max-marginparsing,Yu_learningstructural,SOPBook}, (b) evaluating generalisation bounds in structured prediction \cite{predictingStructuredData, McAllesterK11, JMLR:v17:15-501,SOPBook,RanjbarMW10,RaeSon07,tarlow2012structured,Taskar04learningstructured,LiuSheng2013,KuleszaP07,Finley/Joachims/08a,MeshiSJG10,PSS15,ShevadePSK11,Taskar06structuredprediction}, and (c) performing MAP inference on otherwise tractable models subject to global constraints \cite{LimJK14, Nowozin:2014:ASP:2627999, DBLP:conf/icml/MartinsFASX11}. The latter covers various search problems including the special task of \emph{(diverse) k-best MAP inference} \cite{BatraYGS12, Guzman-RiveraKB13}. Learning with non-decomposable loss functions, in particular, benefits from an efficient inference as all of the theoretical guarantees of training with SSVMs assume exact inference during optimisation \cite{Tsochantaridis05largemargin, Joachims09lcuttingPlane, kelleyjr1960cpm, Lacoste-JulienJSP13, TeoVSL10, SmolaVL07}.

Previous work \cite{ BauerGBMK14, BauerBM17, BauerSM17, OptMPMP} focusing on the problem of loss-augmented inference (use case (a)) has demonstrated how efficient computation can be performed on models with specific global factors. The proposed idea models non-decomposable functions as a kind of multivariate cardinality potentials $\eta(\bfG(\cdot))$ where $\eta \colon \mathbb{R}^P \rightarrow \mathbb{R}_+$ is some function and $\bfG$ denotes sufficient statistics of the global term.
While being able to model popular performance measures, the objective of a corresponding inference problem is rather restricted towards simple interactions between the energy of the core model $F$ and the sufficient statics $\bfG$ according to $F \odot \eta(\bfG)$ where $\odot \colon \mathbb{R} \times \mathbb{R} \rightarrow \mathbb{R}$ is either summation or multiplication operation. Although the same framework can be applied for use case (c) by modelling global constraints via indicator function, it cannot handle a range of other problems in use case (b) which introduce more subtle dependencies between $F$ and $\bfG$.

In this paper, we extend the framework for an efficient exact inference proposed in \cite{BauerSM17} by allowing much finer interactions between the energy of the core model and the sufficient statistics of the global terms.
The extended framework covers all the previous cases
and applies to new problems including evaluation of generalisation bounds in structured learning, which cannot be handled by the previous approach.
At the same time, the generalisation comes with no additional cost preserving all the run-time guarantees. In fact, the resulting performance is identical with that of the previous formulation as the corresponding modifications do not change the computational core idea of previously proposed message passing constrained via auxiliary variables but only affect the final evaluation step (line 8 in Algorithm \ref{algLAMP2}) of the resulting inference algorithm after all the required statistics have been computed. We accordingly adjust the formal statements given in \cite{BauerSM17} to ensure the correctness of the algorithmic procedure for the extended case.
Furthermore, we propose an additional graph transformation technique via node cloning which greatly improves upon the theoretical guarantees on the asymptotic upper bound for the computational complexity.
In particular, previous work only guarantees polynomial run-time in the case where the core model can be represented by a tree-shaped factor graph excluding problems with cyclic dependencies. A corresponding estimation for clique trees (Theorem 2 in \cite{BauerSM17}), however, requires the maximal node degree $\nu$ to be bounded by a graph independent constant which otherwise results in an exponential run-time. Here, we first provide an intuition that $\nu$ tends to take on small values (Proposition \ref{p_nu}) and then present an additional graph transformation which reduces this parameter to a constant $\nu = 3$ (Corollary \ref{cor_1}).
Furthermore, we analyse how the maximal number of states of auxiliary variables $R$, which greatly affects the resulting run-time,
behaves relatively to the graph size (Theorem \ref{theorem3}). 

The rest of the paper is organised as follows. In Section \ref{sec2} we formally introduce a class of problems we are tackling in this paper and present in Section \ref{sec3} a message passing algorithm for finding a corresponding optimal solution. We propose an additional graph transformation technique via node cloning which ensures a polynomial run-time independently of the form of a corresponding clique tree. In Section \ref{sec4} we demonstrate the expressivity of our abstract problem formulation on several examples. For an important use case of loss augmented inference for SSVMs we show in Section \ref{sec5} how to write different dissimilarity measure in a required form as global cardinality potentials. In Section \ref{sec6} we discuss the pervious works and summarise the differences to our approach. We validate the guarantees on the computational time complexity in Section \ref{sec7} followed by a conclusion in Section \ref{sec8}.

\section{Problem Setting}
\label{sec2}
Given an MRF \cite{Bishop:2006:PRM:1162264, Koller:2009:PGM:1795555} over a set of discrete variables, the goal of the maximum a posteriori (MAP) problem is to find a joint variable assignment with the highest probability and is equivalent to the problem of minimising the energy of the model, which describes a corresponding (unnormalised) probability distribution over the variables.
In the context of structured prediction it is equivalent to a problem of maximising a score or compatibility function. To avoid ambiguity, we now refer to the MAP problem as a maximisation of an objective function $F: \mathbb{R}^{M} \to \mathbb{R}$ defined over a set of discrete variables $\bfy = (y_1, ..., y_M)$. More precisely, we associate each function $F$ with an MRF where each variable $y_m$ represents a node in a corresponding graph. Furthermore, we assume without loss of generality that the function $F$ factorises over maximal cliques $\bfy_{C_t}$, $C_t \subseteq \{1, ..., M\}$ of a corresponding MRF according to
\begin{equation}
F({\bfy}) \textstyle = \sum_{t=1}^T f_t(\bfy_{C_{t}}).
\end{equation}
We now use the concept of treewidth\footnote{Informally, the treewidth describes the tree-likeness of a graph, that is, how good the graph structure resembles the form of a tree. In an MRF with no cycles going over the individual cliques the treewidth is equal to the maximal size of a clique minus 1, that is, $\tau = \max_t |C_t| - 1$.} of a graph \cite{Bodlaender93} to define the complexity of a corresponding function with respect to the MAP inference as follows:
\begin{definition}[$\tau$-decomposability] \label{def:TauDecomposability} 
We say that a function $F: \mcD \subseteq \mathbb{R}^M \to \mathbb{R}$ is $\tau$-\emph{decomposable}, 
if the (unnormalised) probability $\exp(F(\bfy))$
factorises over an MRF with a bounded\footnote{The treewidth of a graph is considered to be bounded if it does not depend on the size of the graph. If there is a way of increasing the graph size by replicating individual parts, the treewidth must not be affected by the number of the variables in a resulting graph. One simple example is a Markov chain. Increasing the length of the chain does not affect the treewidth being equal to the Markov order of that chain.} treewidth  $\tau$.
\end{definition}

The treewidth is defined as the minimal width of a graph and as such can be computed algorithmically after transforming a corresponding MRF into a data structure called junction tree or clique tree. Although the problem of constructing a clique tree with a minimal width is NP-hard in general, there are several efficient techniques \cite{Koller:2009:PGM:1795555} which provide good results with a width being close to the treewidth.

In the following, let $M$ be the total number of nodes in an MRF over the variables $\{y_m\}_{m=1}^M$ and $N$ be the maximum number of the possible values each variable $y_m$ can take on.
Provided the maximisation part dominates the time for the creating a clique tree, we get
the following known result \cite{Lauritzen:1990:LCP:84628.85343}:
\begin{proposition}
\label{PropositionDP}
The computational time complexity for maximising a $\tau$-decomposable function is upper bounded by $O(M \cdot N^{\tau+1} )$.
\end{proposition}

The notion of $\tau$-decomposability for real-valued functions naturally extends to mappings with multivariate outputs for which
we now define joint decomposability:
\begin{definition}[Joint $\tau$-decomposability] \label{def:JointTauDecomposability} 
We say two mappings $\bfG: \mcD \subseteq \mathbb{R}^M \to \mathbb{R}^P$ and $\bfG': \mcD \subseteq \mathbb{R}^M \to \mathbb{R}^{P'}$
are \emph{jointly $\tau$-decomposable}, if they factorise over a common MRF with a bounded treewidth $\tau$.
\end{definition}
\noindent Definition \ref{def:JointTauDecomposability} ensures the existence
of a common clique tree with nodes $\{C_t\}_{t=1}^{T}$ and the corresponding potentials $\{\bfg_t, \bfg_t'\}_{t=1}^{T}$ where $\max_t |C_{t}| - 1= \tau$, and 
\begin{align*}
\textstyle
\bfG(\bfy)  \!=\! \sum_{t=1}^{T}  \bfg_t(\bfy_{C_{t}}), \hspace*{5pt}\bfG'(\bfy)  \!=\! \sum_{t=1}^{T} \! {\bfg}_t'(\bfy_{C_{t}}).
\notag
\end{align*}
Note that the individual factor functions are allowed to have less variables in their scope
than in a corresponding clique, that is,  $\text{scope}(\bfg_t), \text{scope}(\bfg'_t) \subseteq C_t$.
\begin{figure*}[t]
\centering
\includegraphics[scale = 1.0]{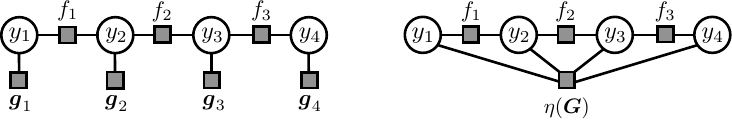}
\caption{Factor graph representation for the margin scaling objective
with a decomposable loss $\bfG$ (on the left), 
and a (non-decomposable) high-order loss $\eta(\bfG)$ (on the right).
}
\label{fig_bauer4}
\end{figure*}
Building on the above definitions we now formally introduce a class of problem instances of MAP inference
for which we later provide an exact message passing algorithm.
\begin{problem} \label{mainproblem}
For $F\colon \mcY \to \mathbb{R}$, $\bfG\colon \mcY \to 
\mathbb{R}^P$, $H \colon \mathbb{R} \times \mathbb{R}^P \rightarrow \mathbb{R}$
with $\mcY \subset \mathbb{R}^M$, $|\mcY| \leqslant N^M$ and $P, M, N \in \mathbb{N}$,
we consider the following discrete optimisation problem:
\begin{equation}
\begin{aligned}
& \underset{\bfy \in \mcY}{\text{maximise}}
& & H(F(\bfy), \bfG(\bfy))
\end{aligned}
\label{Objective2}
\end{equation}
where we assume that:
1) $F$ and $\bfG$ are jointly $\tau$-decomposable,
2) $H$ is non-decreasing in the first argument.
%2) $R$ (as defined in Eq.\ (\ref{E_R})) is polynomial in $M$.
\end{problem}

\noindent In the next section we show multiple examples of practical problems matching the above abstract formulation.
 As our working example we here consider the problem of loss augmented inference within the framework of SSVM. The latter comes with two different formulations called margin and slack scaling  and repeatedly requires solving a combinatorial optimisation problem during training either to compute the subgradient of a corresponding objective function or to find the most violating configuration of the prediction variables with respect to the problem constraints.
For example, for slack scaling formulation, we could define $H(F(\bfy), \bfG(\bfy)) = F(\bfy) \cdot \eta(\bfG(\bfy))$
for some $\eta \colon \mathbb{R}^P \rightarrow \mathbb{R}_+$,
where $F(\bfy) =  \bfw^\T \bfPsi(\bfx, \bfy)$ corresponds to the compatibility given as inner product between a joint feature map $\bfPsi(\bfx, \bfy)$ 
and a vector of trainable weights $\bfw$ (see \cite{Tsochantaridis05largemargin} for more details) and
$\eta(\bfG(\bfy)) = \Delta(\bfy^*, \bfy)$ describes a corresponding loss function for a prediction $\bfy$ and a ground-truth output $\bfy^*$.
In fact, a considerable number of popular loss functions used in structured prediction can be generally represented in this form, that is, as a multivariate cardinality-based potential based on counts of different label statistics.

\section{Exact Inference for Problem \ref{mainproblem}}
\label{sec3}
In this section we derive a polynomial-time message passing algorithm which always finds an optimal solution for Problem \ref{mainproblem}. The corresponding result can be seen as a direct extension of the well-known junction tree algorithm.

\subsection{Algorithmic Core Idea for a Simple Chain Graph}
We begin by giving an intuition why efficient inference is possible for Problem \ref{mainproblem} using our woking example of loss augmented inference for SSVMs. For margin scaling, in case of a linear $\eta$,
the objective $F(\bfy) + \eta(\bfG(\bfy))$ inherits the $\tau$-decomposability directly from $F$ and $\bfG$ and, therefore, can be efficiently maximised 
according to Proposition~\ref{PropositionDP}.

The main source of difficulty for slack scaling lies in
the multiplication operation between $F(\bfy) $ and $\eta(\bfG(\bfy))$,
which results in a fully-connected MRF regardless of the form of the function $\eta$.
Moreover, many popular loss functions used in structured learning require $\eta$ to be non-linear preventing efficient inference even for the margin scaling.
Nevertheless, an efficient inference is possible for a considerable number of practical case as shown below.
Namely, the global
interactions between jointly decomposable $F$ and $\bfG$ can be controlled using auxiliary variables at a polynomial cost.
We now illustrate this on a simple example.

Consider a (Markov) chain of nodes with a $1$-decomposable $F$ and $0$-decomposable $\bfG$ (e.g. Hamming distance).
That is,
\begin{equation}
F(\bfy) = \sum_{m=1}^{M-1} f_m(y_m, y_{m+1}) \hspace*{5pt}\textrm{and}\hspace*{5pt} \bfG(\bfy) = \sum_{m=1}^M  \bfg_m(y_m).
\end{equation}
We aim at maximising an objective
$F(\bfy) \odot \eta(\bfG(\bfy))$ where $\odot \colon \mathbb{R} \times \mathbb{R} \rightarrow \mathbb{R}$ is a placeholder for either summation or multiplication operation.
\begin{figure*}[t]
\centering
\includegraphics[scale = 1.0]{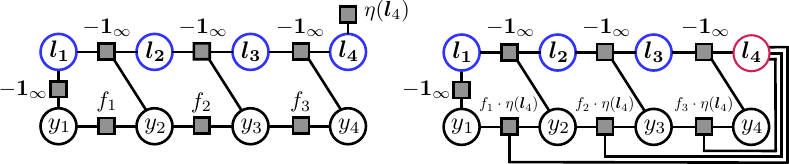}
\caption{Factor graph representation for an augmented objective $Q(\bfy, \bfL)$ for margin scaling (on the left) and for slack scaling (on the right).
The auxiliary variables $\bfL = (\bfl_1, ..., \bfl_4)$
are marked blue (except $\bfl_4$ for slack scaling).
$\bfl_4$ is the hub node.
}
\label{fig_bauer4_}
\end{figure*}
The case for margin scaling with a decomposable loss $\eta(\bfG) = \bfG$ is illustrated by the leftmost factor graph in Figure~\ref{fig_bauer4}.
Here, the corresponding factors $f_m$ and $\bfg_m$ can be folded together
enabling an efficient inference according to Proposition \ref{PropositionDP}.
The nonlinearity of $\eta$, however, results
(in the worst case!)
in a global dependency between all the variable nodes
giving rise to a high-order potential $\eta(\bfG)$ as illustrated
by the rightmost factor graph in Figure~\ref{fig_bauer4}.
In slack scaling, even for a linear $\eta$, after multiplying the individual factors
we can see that the resulting model has an edge for every pair of variables
resulting in a fully-connected graph.
Thus, for the last two cases
an exact inference is infeasible in general.
The key idea is to relax the dense connections in these graphs
by introducing \emph{auxiliary variables}
$\bfL = (\bfl_1, \ldots, \bfl_M)\in \mathbb{R}^{P \times M}$ subject to the constraints
\begin{equation}
\bfl_m = \sum_{k=1}^{m} \bfg_{k}(y_{k}), \hspace*{5pt} m \in \{1, ..., M\}.
\label{eq:DefinitionL}
\end{equation}
More precisely, for $H(F(\bfy), \bfG(\bfy)) = F(\bfy) \odot \eta(\bfG(\bfy))$, Problem \ref{mainproblem}
is equivalent to the following constrained optimisation problem in the sense that both have the same optimal value and  the same set of optimal solutions with respect to $\bfy$:
\begin{equation}
\begin{aligned}
& \underset{\bfy, \bfL}{\text{maximise}}
& & F(\bfy) \odot \eta(\bfl_M )\\
& \text{subject to}
& & \bfl_{m+1} = \bfl_{m} + \bfg_{m+1}(y_{m+1})\hspace*{10pt} \forall m\\
& & & \bfl_{1} = \bfg_1(y_{1})
\end{aligned}
\label{OP1}
\end{equation}
where the new objective involves no global
dependencies , and is $1$-decomposable
if we regard $\eta(\bfl_M)$ as a constant.
We can make the local dependency structure of the new formulation more explicit
by taking the constraints directly into the objective as follows
\begin{equation}
\begin{aligned}
{Q}(\bfy, \bfL) =& F(\bfy) \odot \eta(\bfl_M ) - \mathbb{1}_\infty[\bfl_{1} \ne \bfg_1(y_{1})]\\
 -& \sum_{m=1}^{M-1} \mathbb{1}_\infty[\bfl_{m+1} \ne \bfl_{m} + \bfg_{m+1}(y_{m+1})].
\end{aligned}
\label{eq:AugmentedObjective}
\end{equation}
Here, $\mathbb{1}_\alpha[\cdot]$ denotes the indicator function such that $\mathbb{1}_\alpha[\cdot] = \alpha$ if the argument in $[\cdot]$ is true
and $\mathbb{1}_\alpha[\cdot] = 0$ otherwise.
The indicator functions
rule out the configurations that do not satisfy Eq.\ (\ref{eq:DefinitionL}) when maximisation is performed.
A corresponding factor graph for margin scaling is illustrated by the leftmost graph in Figure~\ref{fig_bauer4_}.
We see that our new augmented objective (\ref{eq:AugmentedObjective})
shows only local dependencies and is, in fact, $2$-decomposable.

Applying the same scheme for slack scaling also yields
a much more sparsely-connected graph (see the rightmost graph in Figure\ \ref{fig_bauer4_}) by forcing the most connections to go through a single node
$\bfl_4$, which we call a \emph{hub} node.
Actually, ${Q}(\bfy, \bfL)$ becomes $2$-decomposable if we fix the value of $\bfl_4$,
which then can be multiplied into the corresponding factors of $F$.
This way
we can effectively reduce the overall treewidth
at the expense of an increased \textbf{polynomial} computation time (compared to the chain without the global factor), provided the maximal number $R$
of different states of each auxiliary variable is polynomially bounded in $M$, the
number of nodes in the original graph. In the context of training SSVMs, for example, the most of the popular loss functions satisfy this condition (see Table \ref{tab:DissilarityMeasure} in Section \ref{sec5} for an overview).

\subsection{Message Passing Algorithm on Clique Trees}
\label{sec:LMP}
The idea presented in the previous section is intuitive and allows for reusing of existing software.
After the corresponding graph transformation due to the introduction of auxiliary variables we can use the standard junction tree algorithm for
graphical models.
Alternatively to the explicit graph transformation we can modify the message passing protocol instead,
which is asymptotically at least one order of magnitude faster.
Therefore, we do not explicitly introduce auxiliary variables as graph nodes
during construction of the clique tree
but use them to condition the message passing rules
as we shall see shortly.
In the following we derive an algorithm for solving an instance of Problem \ref{mainproblem} via message passing
on clique trees.

First, similar to conventional junction tree algorithm, we need to construct a clique tree\footnote{A clique tree is constructed from a corresponding MRF after removing the global term. The corresponding energy is given by the function $F$ according to definition in Problem \ref{mainproblem}.}
for a given set of factors,
which is family preserving and has the running intersection property.
There are two equivalent approaches \cite{Bishop:2006:PRM:1162264, Koller:2009:PGM:1795555}, the first  based on variable elimination
and the second on graph triangulation, upper bounded by $O(M \cdot N^{\tau+1})$.
Assume now that a clique tree with cliques $C_1, ..., C_K$ is given,
where $C_i$ denotes a set of indices of variables contained in the $i$-th clique. A corresponding set of variables is given by $\bfy_{C_i}$.
We denote the clique potentials (or factors) related to the mappings $F$ and $\bfG$ (see Problem \ref{mainproblem}) by $\{f_{C_i}\}_{i = 1}^K$ and $\{\bfg_{C_i}\}_{i = 1}^K$, respectively.
Additionally, we denote by $C_r$ a clique chosen to be the root of the clique tree.
Finally, we use the notation $ne(C_i)$ for the indices of the neighbours of the clique $C_i$.
We can now compute the optimal value of the objective in Problem \ref{mainproblem} as follows.
Starting at the leaves of the clique tree we iteratively send messages toward the root
according to the following message passing protocol.
A clique $C_i$ can send a message to its parent clique $C_j$
if it received all messages from all its other neighbours $C_k$ for $k \in ne(C_i)\setminus \{j\}$.
In that case we say that $C_i$ is \emph{ready}.

For each configuration of the variables $\bfy_{C_i \cap C_j}$ and parameters $\bfl_i \in \mathbb{R}^{P}$
(encoding the state of an auxiliary variable associated with the current clique $C_i$)
a corresponding message from a clique $C_i$ to a clique $C_j$ can be computed
according to the following equation
\begin{equation}
\begin{aligned}
\mu_{C_i \rightarrow C_j}^{\bfl_i}(\bfy_{C_i \cap C_j}) =& \max_{\bfy_{C_i \setminus C_j}, \{\bfl_k\}} f_{C_i}(\bfy_{C_i})\\
+& \sum_{k \in ne(C_i)\setminus\{j\}} \mu_{C_k \rightarrow C_i}^{\bfl_k}(\bfy_{C_k \cap C_i})
\end{aligned}
\label{E_mp_}
\end{equation}
where we maximise over all configurations of the variables $\bfy_{C_i \setminus C_j}$ and over all parameters $\{\bfl_k\} = \{ \bfl_k \colon k \in ne(C_i)\setminus\{j\}\}$
subject to the following constraint
\begin{equation}
\label{eq_aux}
\sum_{k \in ne(C_i)\setminus\{j\}} \bfl_k = \bfl_i - \bfg_{C_i}(\bfy_{C_i}).
\end{equation}
That is, each clique $C_i$ is assigned with exactly one (multivariate) auxiliary variable $\bfl_i$
and the range of possible values $\bfl_i$ can take on is implicitly defined by the equation (\ref{eq_aux}).
After resolving the recursion in the above equation, we can see that the variable $\bfl_i$ corresponds
to a sum of the potentials $\bfg_{C_k}(\bfy_{C_k})$ for each previously processed clique $C_k$ in a subtree of the graph
of which $C_i$ forms the root. We refer to the equation (\ref{eq:DefinitionL}) in the previous subsection for comparison.

\begin{algorithm}[b]
\caption{Inference on a Clique Tree}
\label{algLAMP2}
\small
%\hspace*{\algorithmicindent} \textbf{Input:} clique tree; \textbf{Output:} $\bfy^*$
\begin{algorithmic}[1]
	\REQUIRE clique tree $\{C_i\}_{i}$; \textbf{Output:} optimal assignment $\bfy^*$
	%\ENSURE Output...
	%\STATE{Compute the initial messages for leaf nodes according to Eq.\ (\ref{E_mp01}) and (\ref{E_mp02})}
	\WHILE{root clique $C_r$ did not receive all messages}
		\IF{a clique $C_i$ is ready}
			\FOR{all $\bfy_{C_i \cap C_j}$ and $\bfl$}
				\STATE{send a message $\mu_{C_i \rightarrow C_j}^{\bfl_i}(\bfy_{C_i \cap C_j})$ to a parent clique $C_j$ according to Eq.\ (\ref{E_mp_});
					     save the maximising arguments $\lambda_{C_i \rightarrow C_j}^{\bfl_i}(\bfy_{C_i \cap C_j}) := [\bfy_{C_i \setminus C_j}^*; \{\bfl_k\}^*]$}
			\ENDFOR
		\ENDIF
	\ENDWHILE
	\STATE $\bfl^* \leftarrow \argmax_{\bfl}\hspace*{1pt}H(\mu(\bfl), \bfl)$, \hspace*{5pt} where $\mu(\bfl)$ is defined by Eq. (\ref{beliefs_})
	\STATE Let $\bfy_{C_r}^*$ be a maximising argument for $\mu(\bfl^*)$ in Eq.\ (\ref{beliefs_}); 
		     starting with values $\bfl^*$ and $\bfy_{C_r}^*$ recursively reconstruct an optimal configuration $\bfy^*$ from $\lambda$ according to the formula in Eq. (\ref{E_mp_}).
\end{algorithmic}
\end{algorithm}

\begin{figure*}[t]
\centering
\includegraphics[scale = 0.9]{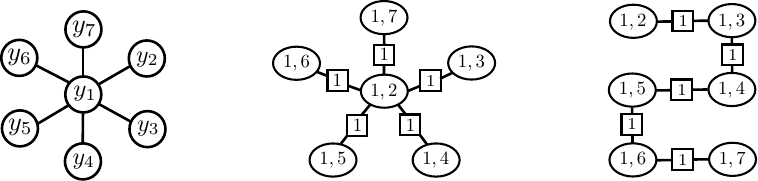}
\caption{Illustration of an extreme example where $\nu$ can be linear in the graph size. The leftmost graph represents an MRF with $M=7$ variables and treewidth $\tau = 1$. The graph in the middle shows a valid clique tree for the MRF on the left where the clique $\{y_1, y_2\}$ has $M-1$ neighbours. That is, $\nu$ is linear in the graph size for that clique tree. The rightmost graph represents another clique tree which has a chain form where $\nu = \tau+1 = 2$. The squared nodes denote here the corresponding sepsets.
}
\label{fig_example_nu}
\end{figure*}

The algorithm terminates if the designated root clique $C_r$ received all messages from its neighbours.
We then compute the values
\begin{equation}\label{beliefs_}
\mu({\bfl}) = \max_{\bfy_{C_r}, \{\bfl_k\}} f_{C_r}(\bfy_{C_r}) + \sum_{k \in ne(C_r)} \mu_{C_k \rightarrow C_r}^{\bfl_k}(\bfy_{C_k \cap C_r})
\end{equation}
maximising over all configurations of $\bfy_{C_r}$ and $\{\bfl_k\} = \{ \bfl_k \colon k \in ne(C_r)\}$ subject to the constraint $\sum_{k} \bfl_k = \bfl - \bfg_{C_r}(\bfy_{C_r})$,
which we use to get the optimal value $p^*$ of Problem~\ref{mainproblem} according to
\begin{equation}\label{solution_}
p^* = \max_{\bfl}\hspace*{1pt} H(\mu(\bfl), \bfl).
\end{equation}
A corresponding optimal solution of Problem \ref{mainproblem} can be obtained by backtracking the additional variables $\lambda$
saving optimal decisions in intermediate steps.
The complete algorithm is summarised in
Algorithm \ref{algLAMP2} supported by the following theorem\footnote{The theorem refers to a more general target objective defined in Problem \ref{mainproblem} and should replace a corresponding statement in Theorem 2 in \cite{BauerSM17}.}, for which we provide a proof in the appendix \ref{sec:Proof2}.
\begin{theorem}
\label{theorem2}
Algorithm \ref{algLAMP2} always finds an optimal solution of Problem \ref{mainproblem}.
The computational complexity is of the  order $O(M \cdot N^{\tau+1} \cdot {R}^{{\nu}-1})$
where $\nu$ is defined as the maximal number of neighbours
of a node in a corresponding clique tree.
\end{theorem}

Besides the treewidth $\tau$, the value of the parameter $\nu$ also appears to be crucial for the resulting running time of Algorithm \ref{algLAMP2} since the corresponding complexity is also exponential in $\nu$.
The following proposition suggests that among all possible cluster graphs for a given MRF there always exist a clique tree for which $\nu$ tends to take on small values (provided $\tau$ is small) and effectively does not depend on the size of a corresponding MRF. We provide a proof in the appendix \ref{sec:Proof_prop}.
\begin{proposition}
\label{p_nu}
For any MRF with treewidth $\tau$, there is a clique tree, for which the maximal number of neighbors $\nu$ is upper bounded according to $\nu \leqslant 2^{\tau+2}-4$.
\end{proposition}
To support the above proposition we consider the following extreme example illustrated in Figure \ref{fig_example_nu}. We are given an MRF with a star-like shape (on the left) having $M = 7$ variables and treewidth $\tau = 1$. One valid clique tree for this MRF is shown in the middle. In particular, the clique containing the variables $y_1, y_2$ has $\nu = M-1$ neighbours. Therefore, running Algorithm \ref{algLAMP2} on that clique tree results in a computational time exponential in the graph size $M$. However, it is easy to modify that clique tree to have a small number of neighbours for each node (shown on the right) upper bounded by $\nu = \tau+1 = 2$.

\begin{figure*}[t]
\centering
\includegraphics[scale = 0.7]{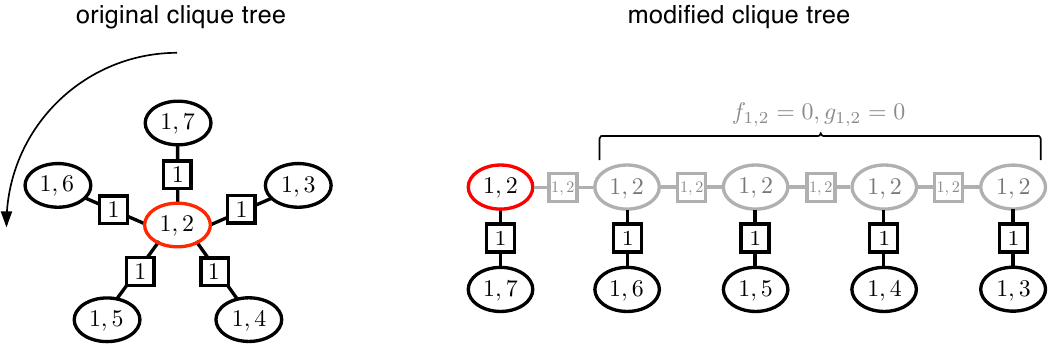}
\caption{Illustration of a modification procedure to reduce the maximal number of neighbours $\nu$ for each cluster node in a given clique tree. A graph on the left represents an original clique tree. The only node having more than three neighbours is marked red. We clone this cluster node multiple times so that each clone only carries one of the original neighbours and connect the clones by a chain which preserves the running intersection property. The arrow at the left graph indicates an (arbitrarily chosen) order of processing the neighbour nodes. The graph resulting after this transformation is shown on the right. The clones in the new graph are marked gray. To ensure that the new cluster graph describes the same set of potentials we set the potentials for copies of each cloned cluster node $C_i$ to zero: $f_{C_i}(\bfy_{C_i}) = 0$ and $\bfg_{C_i}(\bfy_{C_i}) = 0$. This procedure reduces the $\nu$-parameter to a constant ($\nu = 3$) significantly reducing the computational cost.
}
\label{fig_bauer4_1}
\end{figure*}

Although Proposition \ref{p_nu} assures an existence of a clique tree with a small $\nu$, the actual upper bound on $\nu$ is still very pessimistic (exponential in the treewidth). In fact, by allowing a simple graph modification we can always reduce the $\nu$-parameter to a small constant ($\nu = 3$).
Namely, we can clone each cluster node with more than three neighbours multiple times so that each clone only carries one of the original neighbours and connect the clones by a chain which preserves the running intersection property. To ensure that the new cluster graph describes the same set of potentials we set the potentials for each copy of a cluster node $C_i$ to zero: $f_{C_i}(\bfy_{C_i}) = 0$ and $\bfg_{C_i}(\bfy_{C_i}) = 0$. The whole modification procedure is illustrated in Figure \ref{fig_bauer4_1}. We summarise this result in the following corollary.
\begin{corollary}
\label{cor_1}
Provided a given clique tree is modified according to the presented procedure for reducing the number of neighbours for each cluster node, 
the overall computational complexity of running Algorithm \ref{algLAMP2} (including time for graph modification) is of the order $O(M \cdot N^{\tau+1} \cdot {R}^2)$.
\end{corollary}

Note that in case where the corresponding clique tree is a chain the resulting complexity reduces to $O(M \cdot N^{\tau+1} \cdot {R})$.
At this point we would like to give an alternative view on the computational complexity in that case (with $\tau = 1$, $R \sim M^2$) which shows the connection to the conventional junction tree algorithm. Namely, the constraint message passing algorithm (Algorithm \ref{algLAMP2}) can be seen as a conventional message passing on a clique tree (for the mapping $F$ in Problem \ref{mainproblem}) without auxiliary variables, but where the size of the state space for each variable $y_i$ is increased from $N$ to $N \cdot M$. Then Proposition \ref{PropositionDP} guarantees an exact inference in time of the order $O(M \cdot (N \cdot M)^{\tau+1})$.
The summation constraints with respect to the auxiliary variables can be ensured by extending the corresponding potential functions $f_{C_i}$ to take on $-\infty$ forbidding inconsistent state transitions between individual variables. The same observation holds also for message passing on factor graphs.
To summarise, we can remove the global dependencies (imposed by the mapping $H$ in Problem \ref{mainproblem}) reducing the overall treewidth by introducing auxiliary variables, but pay a price that the size of the label space becomes a function of the graph size ($R$ is usually dependent on $M$).

We complete our discussion by analysing the relation between the maximal number of states (of the auxiliary variables) $R$
and the number of variables $M$ in the original MRF.
In the worst case, $R$ can be exponential in the graph size $M$.
This happens, for example, if the values, the individual factors $\bfg_{C_k}$ can take on, are scattered in a very long range
which grows much faster relative to the graph size.
For practical cases, however, we can assume
the individual factor functions $\bfg_{C_k}$ to take values in an integer interval, which is either fixed
or grows polynomially with the graph size.
In that case, $R$ is always a polynomial in $M$ rendering the overall complexity of Algorithm \ref{algLAMP2} a polynomial in the graph size
as we show for several examples in the experimental section \ref{sec7}.
We summarise this fact in the following theorem. A corresponding proof is given in the appendix \ref{sec:Proof4}.
\begin{theorem}
\label{theorem3}
Consider an instance of Problem \ref{mainproblem} given by a clique tree with $M$ variables.
Let $T \in \mathbb{N}$ be a number growing polynomially with $M$.
Provided each factor $\bfg_{C_k}$ in a decomposition of $\bfG$ assumes values from a discrete set of integers $[-T, T] \cap \mathbb{Z}$,
the number $R$ grows polynomially with $M$ according to $R \sim T \cdot M$.
\end{theorem}

\section{General Use Cases}
\label{sec4}
In this section we demonstrate the expressivity of Problem \ref{mainproblem} by showing a few different examples.

\subsection{Loss Augmented Inference with High Order Loss Functions}
\label{sec4_a}
As already mentioned, Problem \ref{mainproblem} covers as a special case the task of loss augmented inference 
(for margin and slack scaling) within the framework of SSVM \cite{Tsochantaridis05largemargin, BauerBM17}.
Namely, for the generic representation given in (\ref{Objective2}) we can define $F(\bfy) = \bfw^\T \bfPsi(\bfx, \bfy) + \text{const.}$
and $\eta(\bfG(\bfy)) = \Delta(\bfy^*, \bfy)$ for a suitable $\eta \colon \mathbb{R}^P \rightarrow \mathbb{R}_+$.
Here $\bfPsi \colon \mcX \times \mcY \rightarrow \mathbb{R}^d, d \in \mathbb{N}$ denotes a joint feature map on an input-output pair $(\bfx, \bfy)$,
$\bfw \in \mathbb{R}^d$ is a trainable weight vector,
and $\Delta: \mcY \times \mcY \rightarrow \mathbb{R}$ is a dissimilarity measure between a prediction $\bfy$ and a true output $\bfy^*$.
Given this notation our target objective can be written as follows
\begin{equation}
H(F(\bfy), \bfG(\bfy)) = F(\bfy) \odot \eta(\bfG(\bfy)), \hspace*{5pt} \odot \in \{+,\cdot\}.
\end{equation}
We note that a considerable number of non-decomposable (or high order) loss functions in structured prediction can be represented as a multivariate cardinality-based potential $\eta(\bfG(\cdot))$ where the mapping $\bfG$ encodes the label statistics, e.g., the number of true or false positives with respect to the ground truth. Furthermore, the maximal number of states $R$ for the corresponding auxiliary variables related to $\bfG$ is polynomially bounded in the the number of variables $M$.
See Table \ref{tab:DissilarityMeasure} in Section \ref{sec5} for an overview of existing loss functions. For the specific case of a chain graph with $F_{\beta}$-loss the corresponding complexity $O(M^3 \cdot N^2)$ is cubic in the graph size.

\subsection{Evaluating Generalisation Bounds in Structured Prediction}
\label{sec4_b}
Generalisation bounds can give useful theoretical insights in behaviour and stability of a learning algorithm
by upper bounding the expected loss or the risk of a prediction function.
Evaluating such a bound could provide certain guarantees how a system trained
on some finite data will perform in the future on the unseen examples.
Unlike in the standard regression or classification tasks with univariate real-valued outputs,
in structured prediction, evaluating generalisation bounds requires
to solve a complex combinatorial optimisation problem limiting its use in practice.
In the following we demonstrate how the presented algorithmic idea
can be used
to evaluate PAC-Bayesian Generalisation bounds for max-margin structured prediction.
As a working example we consider the following Generalisation theorem stated in \cite{predictingStructuredData}:
\begin{theorem}
\label{T_GB}
Assume that $0 \leqslant \Delta(\bfy^*, \bfy) \leqslant 1$. With probability at least $1 - \delta$
over the draw of the training set $\mcS = \{(\bfx_1, \bfy_1), ..., (\bfx_1, \bfy_n)\}$ of size $n \in \mathbb{N}$,
the following holds simultaneously for all weight vectors $\bfw$:
\begin{equation}
\scriptsize
\begin{aligned}
\mathbf{E}_{(\bfx, \bfy)\sim \rho}[\Delta(\bfy, h_{\bfw}(\bfx))] \leqslant
\frac{\|\bfw\|^2}{n} + \sqrt{\frac{\|\bfw\|^2 \ln (\frac{2dn}{\|\bfw\|^2}) + \ln(\frac{n}{\delta})}{2(n-1)}} \\
+ \frac{1}{n}\sum_{i=1}^n \max_{\hat{\bfy}} \mathbb{1}_{1}\left[  \bfw^\T (\bfPsi(\bfx_i, \bfy_i) - \bfPsi(\bfx_i, \hat{\bfy})) \leqslant \Delta_{\textrm{HD}}(\bfy_i, \hat{\bfy})\right] \cdot \Delta(\bfy_i, \hat{\bfy})
\end{aligned}
\label{E_GB}
\end{equation}
where $h_{\bfw}(\bfx) = \underset{\hat{\bfy}}{\argmax}\hspace*{3pt} \bfw^\T \bfPsi(\bfx, \hat{\bfy})$ denotes a corresponding prediction function.
\end{theorem}
\noindent Evaluating the second term of the right hand side of the inequality (\ref{E_GB})
involves for each data point $(\bfx, \bfy^*)$ a maximisation over $\bfy \in \mcY$ according to
\begin{equation*}
%\small
\max_{\bfy \in \mcY} \mathbb{1}_{1}\left[  \bfw^\T \left(\bfPsi(\bfx, \bfy^*) - \bfPsi(\bfx, \bfy)\right) \leqslant \Delta_{\textrm{HD}}(\bfy^*, \bfy)\right] \cdot \Delta(\bfy^*, \bfy).
\end{equation*}
We now show that this maximisation term is an instance of Problem \ref{mainproblem}.
More precisely, we consider an example with $\tau > 1$ and $F_1$-loss (see Table \ref{tab:DissilarityMeasure}) and define $F(\bfy) = \bfw^\T \bfPsi(\bfx, \bfy)$, and $\eta(\bfG(\bfy)) = \Delta_{F_1}(\bfy^*, \bfy)$, and set $H(F(\bfy),\bfG(\bfy))$ equal to
\begin{equation*}
\small
\mathbb{1}_1\left[\bfw^\T \bfPsi(\bfx, \bfy^*) - F(\bfy) \leqslant |\bfy^*| - G_1(\bfy)+G_2(\bfy)\right]\cdot\eta(\bfG(\bfy))
\end{equation*}
where we use $\Delta_{\textrm{HD}}(\bfy^*, \bfy) = FP + FN$, $FN = |\bfy^*| - TP$,
which removes the need of additional auxiliary variables for Hamming distance reducing the resulting computational cost. Here, $TP$, $FP$, and $FN$ denote
the numbers of true positives, false positives, and false negatives, respectively.
$|\bfy^*|$ denotes the size of the output $\bfy^*$.
Both $|\bfy^*|$ and $\bfw^\T \bfPsi(\bfx, \bfy^*)$ are constant with respect to the maximisation over $\bfy$.
Note also that $H$ is non-decreasing in $F(\bfy)$.
Furthermore, the number of states of the auxiliary variables is upper bounded by $R = M^2$. Therefore, the computational complexity of Algorithm \ref{algLAMP2} here is given by $O(M^{5} \cdot N^{\tau+1})$.

As a final remark we note that training an SSVM corresponds to solving a convex problem but is
not consistent. It fails to converge to the optimal predictor even in the limit of infinite training data (see \cite{Nowozin:2014:ASP:2627999} for more details).
However, minimising the (non-convex) generalisation bound is consistent.
Providing an effective evaluation tool, Algorithm \ref{algLAMP2} potentially could be used for development of new training algorithms based on direct minimisation of such bounds.

\subsection{Globally Constrained MAP Inference}
\label{sec4_c}
Another common use case is performing MAP inference on a model subject to additional constraints on the variables or the range of the corresponding objective.

Note that from a technical perspective, the problem of MAP inference subject to some global constraints on the statistics $\bfG(\bfy)$
is equivalent to the MAP problem augmented with a global
cardinality-based potential $\eta(\bfG(\bfy))$. Namely, we can define $\eta$ as an
indicator function $\mathbb{1}_{-\infty}[\cdot]$, which returns $-\infty$ if the corresponding constraint on $\bfG(\bfy)$ is violated.
Furthermore, the form of $\eta$ does not affect the message passing of the presented algorithm.
We can always check the validity of a corresponding constraint after all the necessary statistics have been
computed.

\subsubsection*{Constraints on Label Counts}
As a simple example consider the binary sequence tagging experiment.
That is, every output $\bfy \in \mcY$ is a sequence and each site in the sequence can be either 0 or 1. Given some prior information on the number $b$ of positive labels
we could improve the quality of the results by imposing a corresponding constraint on the outputs:

\begin{equation}
\underset{\bfy \in \mcY}{\text{maximise  }}  \bfw^\T \bfPsi(\bfx, \bfy) \text{     subject to  } \sum_{i = 1}^M y_i = b
%\label{OP1}
\end{equation}

\noindent We can write this as an instance of Problem \ref{mainproblem} by setting
$F(\bfy) = \bfw^\T \bfPsi(\bfx, \bfy)$, and $G(\bfy) = \sum_{i=1}^M y_i$, and
\begin{equation}
H(F(\bfy), G(\bfy)) = F(\bfy) + \mathbb{1}_{-\infty}[G(\bfy) \neq b].
\end{equation}
Since all the variables in $\bfy$ are binary, the number of states $R$ of the corresponding auxiliary variables $l_m = \sum_{i = 1}^m y_i$ is upper bounded by $M$. Also, because the output graph is a sequence we have $\tau=1$, $\nu = 2$. Therefore, the computational complexity of Algorithm \ref{algLAMP2} here is of the order $O(M^{2} \cdot N^{2})$.

\subsubsection*{Constraints on Objective Value}
We continue with binary sequence tagging example (with pairwise interactions).
To enforce constraints on the score to be in a specific range as in
\begin{equation}
\begin{aligned}
& \underset{\bfy \in \mcY}{\text{maximise}}
& & \bfw^\T \bfPsi(\bfx, \bfy) \\
& \text{subject to}
& & a \leqslant \bfw^\T \bfPsi(\bfx, \bfy) \leqslant b
\end{aligned}
%\label{OP1}
\end{equation}
we first rewrite the prediction function in terms of its sufficient statistics according to
\begin{equation}
\begin{aligned}
\bfw^\T \bfPsi(\bfx, \bfy) =& \sum_{o, s} w_{o,s} \underbrace{\sum_{t = 1}^M \mathbb{1}_1[x_t = o \land y_t = s]}_{=: G_{o,s}(\bfy)}\\
+& \sum_{s_1, s_2} w_{s_1,s_2} \underbrace{\sum_{t = 2}^M \mathbb{1}_1[y_{t-1} = s_1 \land y_t = s_2]}_{=: G_{s_1,s_2}(\bfy)},
\end{aligned}
\end{equation}
$\bfw = (..., w_{o, s}, ..., w_{s_1, s_2}, ...)$, 
$\bfG = (..., G_{o, s}, ..., G_{s_1, s_2}, ...)$,
and define
\begin{equation}
H(F(\bfy), \bfG(\bfy)) = F(\bfy) + \mathbb{1}_{-\infty}[\bfw^\T \bfG(\bfy) \notin [a, b]].
\end{equation}
Note that $\bfG$ contains all the sufficient statistics of $F$ such that $F(\bfy) = \bfw^\T \bfG(\bfy)$.
Here we could replace $a \leqslant \bfw^\T \bfPsi(\bfy) \leqslant b$
by any (non linear) constraint on the sufficient statistics of the joint feature map $\bfPsi(\bfy)$.

The corresponding computational complexity can be derived by considering
an urn problem, one with $D\cdot N$ and one with $N^2$ distinguishable urns and $M$
indistinguishable balls. Here, $D$ denotes the size of the dictionary for the observations $x_t$ in the input sequence $\bfx$.
Note that the dictionary of the input symbols can be large with respect to other problem parameters. However, we can reduce $D$ to the size of the vocabulary part only occurring in the current input $\bfx$.
The first urn problem corresponds to the unary observation-state statistics $G_{o,s}(\bfy)$ and the second to the pairwise statistics for the state transition $G_{s_1,s_2}(\bfy)$.
The resulting number of possible distributions of balls over the urns is given by
\begin{equation}
\underbrace{\binom{M+D\cdot N-1}{M}}_{\leqslant M^{D\cdot N}} \cdot \underbrace{\binom{M+N^2-1}{M}}_{\leqslant M^{N^2}} \leqslant \underbrace{M^{D\cdot N + N^2}}_{= R}. 
\end{equation}
Although the resulting complexity (due to $\nu = 2$) being $O(M^{D\cdot N + N^2+1}\cdot N^{2})$ is still a polynomial in the number of variables $M$,
the degree is quite high such that we can consider only  short sequences.
For practical use we recommend the efficient approximation framework of Lagrangian relaxation and Dual Decomposition \cite{KomodakisPT11,RushC14,Bauer2019}.

\subsubsection*{Constraints on Search Space}
The constraints on the search space can be different from the constraints we can impose on the label counts. For example, we might want to exclude a set of $K$ complete outputs $\{\bfy^{1}, ..., \bfy^{K}\}$ from the feasible set $\mcY$ by using an exclusion potential 
$\mathbb{1}_{-\infty}[\bfy \in \{\bfy^{1}, ..., \bfy^{K}\}]$.

For simplicity, we again consider a sequence tagging example with pairwise dependencies.
Given a set of $K$ patterns to exclude we can introduce auxiliary variables $\bfl_m \in \{0,1\}^K$ where for each pattern $\bfy^k$ we have a constraint\footnote{More precisely, we modify the message computation in (\ref{E_mp_})
with respect to the auxiliary variables
by replacing the corresponding constraints $(\bfl_m)_k = (\bfl_{m-1})_k + \mathbb{1}_1[y_m^k \neq y_m]$
in the maximisation over $\{\bfl_m\}$ by the constraints
$(\bfl_m)_k = \max \{\mathbb{1}_1[y_m^k \neq y_m], (\bfl_{m-1})_k\}$.}
%That is, the $+$-operator has now been replaced by the $\max$-operator. However,
%the corresponding objective still factorizes in the same way such that at the final stage we have $\bfG(\bfy) = \bfl_M$.
$(\bfl_m)_k = \max \{\mathbb{1}_1[y_m^k \neq y_m], (\bfl_{m-1})_k\}$. Therefore, the maximal number of states for $\bfl_m$ is given by $R = 2^{K}$. The resulting computational complexity for finding an optimal solution over $\bfy \in \mcY \setminus \{\bfy^1, ..., \bfy^K\}$
is of the order $O(2^{K} \cdot M \cdot N^{2})$.

A related problem is finding a \emph{diverse} $K$-best solution. Here, the goal is to produce best solutions which are sufficiently different from each other according to some diversity function, e. g., a loss function like Hamming distance $\Delta_{\text{HD}}$.
More precisely,
after computing the MAP solution $\bfy^1$ we
compute the second best (diverse) output $\bfy^2$ with $\Delta_{\text{HD}}(\bfy^1, \bfy^2) \geqslant m_1$. For the third best solution we then require $\Delta_{\text{HD}}(\bfy^1, \bfy^3) \geqslant m_2$ and $\Delta_{\text{HD}}(\bfy^2, \bfy^3) \geqslant m_2$
and so on.
That is, we search for an optimal output $\bfy^{K}$ such that $\Delta_{\text{HD}}(\bfy^{k}, \bfy^{K}) \geqslant m_{K-1}$,
$m_k \in \mathbb{N}$ for all $k \in \{1, ..., K-1\}$.

For this purpose, we define auxiliary variables $\bfl_m \in \{0, ..., M\}^{K-1}$ where for each pattern $\bfy^k$ we have a constraint $(\bfl_m)_k = (\bfl_{m-1})_k + \mathbb{1}_1[y_m^k \neq y_m]$
computing the Hamming distance of a solution $\bfy$ with respect to the pattern $\bfy^k$. Therefore, we can define
\begin{equation}
\begin{aligned}
& H(F(\bfy), G(\bfy)) =\\
& F(\bfy) - \mathbb{1}_{\infty}[G_k(\bfy) \geqslant m_{k} \text{ for } k \in \{1, ..., K-1\}]
\end{aligned}
\end{equation}
where $\bfG = (G_1, ..., G_{K-1})$ and
at the final stage (due to $\bfG(\bfy) = \bfl_M$) we have all the necessary information to evaluate the constraints with respect to the diversity function (here Hamming distance).
The maximal number of states $R$ for the auxiliary variables is upper bounded by $M^{K-1}$. Therefore, the resulting running time is of the order $O(M^{K} \cdot N^{\tau+1})$.

Finally, we note that the concept of diverse $K$-best solutions can also be used during the training of SSVMs to speed up the convergence of a corresponding algorithm by generating diverse cutting planes or subgradients as described in \cite{Guzman-RiveraKB13}.
An appealing property of Algorithm \ref{algLAMP2} is that we get some part of the necessary information for free as a side-effect of the message passing.

\section{Compact Representation of Loss Functions}
\label{sec5}
\begin{sidewaystable*}
\centering
%\small
\large
%\Large
{
\renewcommand{\arraystretch}{1.35}
\begin{tabularx}{0.805\textheight}{|l|c|c|c|} 
\hline
Loss Function & $\bfG(\bfy)$ & $\eta(\bfG(\cdot))$ & $R$\\
%\hline
\hline
Zero-one loss $\Delta_{\textrm{0/1}}$ &  $(\textit{TP},\mathbb{1}_1[\textit{FP} > 0])$ & $\mathbb{1}_1[\max\{M-G_1, G_2\}>0]$ & $2M$\\[5pt]
Hamming distance $\Delta_{\textit{HD}}$ &  $\sum_{t = 1}^{M} \mathbb{1}_1[y_t^* \neq y_t]$ & $G$ & $M$\\[5pt]
Hamming loss $\Delta_{\textit{HL}}$ &  $\sum_{t = 1}^{M} \mathbb{1}_1[y_t^* \neq y_t]$ &$G/M$ & $M$\\[5pt]
Weighted Hamming distance $\Delta_{\textit{WHD}}$ &  $\{\#(s_1, s_2)\}_{s_1,s_2 \in \{1, ..., N\}}$ & $\sum_{s_1, s_2} \text{weight}(s_1, s_2) \cdot G_{s_1,s_2}$ & $M^{N^2}$\\[5pt]
False positives number $\Delta_{\textit{\#FP}}$ & $\textit{FP}$ & $G$ & $M$\\[5pt]
Recall $\Delta_{\textit{R}}$ & $\textit{TP}$ & $1 - G/|\bfy^*|$ & $M$\\[5pt]
Precision $\Delta_{\textit{P}}$ & $(\textit{TP},\textit{FP})$ & $1 - \frac{G_1}{G_1+G_2}$ & $M^2$\\[5pt]
$F_{\beta}$-loss $\Delta_{F_{\beta}}$ & $(\textit{TP},\textit{FP})$ & $1 - \frac{(1+\beta^2) \cdot G_1}{\beta^2 \cdot |\bfy^*| + G_1 + G_2}$ & $M^2$\\[5pt]
Intersection over union $\Delta_{\cap/{\cup}}$ & $(\textit{TP},\textit{FP})$ & $1 - \frac{G_1}{|\bfy^*| + G_2}$ & $M^2$\\[5pt]

Label-count loss $\Delta_{\textit{LC}}$ &  $\sum_{t = 1}^{M} y_t$ &  $\left|G-\sum_{t=1}^M y_t^*\right|$ & $M$\\[5pt]

Crossing brackets number $\Delta_{\textit{\#CB}}$ & $\textit{\#CB}$ & $G$ & $M$\\[5pt]
Crossing brackets rate $\Delta_{\textit{CBR}}$ & $(\textit{\#CB},|\bfy|)$ & $G_1/G_2$ & $M^2$\\[5pt]
BLEU $\Delta_{\textit{BLEU}}$ & $(\textit{TP}_1,\textit{FP}_1, ..., \textit{TP}_K,\textit{FP}_K)$ &  $1 - \textit{BP}(\cdot) \cdot \exp \left(\frac{1}{K}\sum_{k = 1}^K \log p_k\right)$ & $M^{2K}$\\[5pt]
ROUGE-K $\Delta_{\textit{ROUGE-K}}$ & $\{\text{count}(k,X)\}_{k \in \text{grams}(Ref)}$ &  $1 - \frac{\sum_{S \in Ref} \sum_{k \in \text{grams}(S)} \min\{\text{count}(k, S), G_{k}\}}{\sum_{S \in Ref} \sum_{k \in \text{grams}(S)} \text{count}(k,S)}$ & $M^{D}$\\[5pt]

ROUGE-LCS $\Delta_{\textit{ROUGE-LCS}}$ & $\{\textit{LCS}(X,S)\}_{S \in Ref}$ &  $1 - \frac{1}{|Ref|}\sum_{S \in Ref} \frac{(1+\beta^2)P(G_S)\cdot R(G_S)}{\beta^2P(G_S)+R(G_S)}$ & $M^{2|Ref|}$\\[5pt]

%Structured AUC & $\Delta_{\textit{SAUCE}}$ & $(\textit{TP}_1,\textit{FP}_1, ..., \textit{TP}_N,\textit{FP}_N)$ &  $1 - \textit{BP}(\cdot) \cdot \exp \left(\frac{1}{N}\sum_{n = 1}^N \log p_n\right)$ & $M^{2N}$ & \xmark \\[5pt]
\hline
\end{tabularx}
}
\normalsize
\caption{Compact representation of popular dissimilarity measures based on the corresponding sufficient statistics $\bfG(\cdot)$.}
%\caption{Dissimilarity measures $\Delta(y^*, y) = \eta(\bfG(y))$.}
\label{tab:DissilarityMeasure}
\end{sidewaystable*}

We now further advance the task of loss augmented inference (see Section \ref{sec4_a}) by presenting a list of popular dissimilarity measures which our algorithm can handle, summarised in Table \ref{tab:DissilarityMeasure}.
The measures are given in a compact representation  $\Delta(\bfy^*, \bfy) = \eta(\bfG(\bfy))$ based on the corresponding sufficient statistics encoded as a mapping $\bfG$. Column 2 and 3 show the form of $\bfG(\cdot)$
and $\eta$, respectively. Column 4 gives an upper bound $R$
on the number of possible values of auxiliary variables affecting the resulting running time  of Algorithm \ref{algLAMP2} (see Corollary \ref{cor_1}).

Here, $|\bfy| = M$ denotes the number of nodes of the output $\bfy$.
$\textit{TP}, \textit{FP}$, and $\textit{FN}$ are the number of 
true positives, false positives, and false negatives, respectively. 
The number of true positives for a prediction $\bfy$
and a true output $\bfy^*$ is defined as the number of common nodes with the same label. The number of false positives is given by the number of nodes which are present in the output $\bfy$ but missing (or having other label) in the true output $\bfy^*$. Similarly, the number of false negatives corresponds to the number of nodes present in $\bfy^*$ but missing (or having other label) in $\bfy$. In particular, it holds $|\bfy^*| = \textit{TP} + \textit{FN}$.

We can see in Table \ref{tab:DissilarityMeasure} that each element of $\bfG(\cdot)$ is a sum of binary variables
significantly reducing the image size of mapping $\bfG(\cdot)$,
despite the exponential variety of the output space $\mcY$.
Due to this fact the image size grows only polynomially with the size of the outputs $\bfy \in \mcY$, and the number $R$ provides an upper bound on the image size of $\bfG(\cdot)$.\\

\noindent \textbf{Zero-One Loss} ($\Delta_{\textrm{0/1}}$)\\
\noindent This loss function takes on binary values $\{0,1\}$ and is the most uninformative since it requires a prediction
to match the ground truth to $100$ $\%$ and  gives no partial quantification
of the prediction quality in the opposite case.
Technically, this measure is not decomposable since it requires the numbers $\textit{FP}$ and $\textit{FN}$ to be evaluated via
\begin{equation}
\Delta_{\textrm{0/1}}(\bfy^*, \bfy) = \mathbb{1}_1[\max\{\textit{FP}, \textit{FN}\}>0].
\end{equation}
Sometimes\footnote{For example, if the outputs $\bfy \in \mcY$ are sets with no ordering indication of the individual set elements, we need to know the whole set $\bfy$ in order to be able to compute $\textit{FN}$. Therefore, computing $\textit{FN}$ from a partially constructed output is not possible.} we cannot compute $\textit{FN}$ (unlike $\textit{FP}$) from the individual nodes of a prediction.
Instead, we can count $\textit{TP}$ and compute $\textit{FN}$ using the relationship $|\bfy^*| = \textit{TP} + \textit{FN}$. We note, however, that in case of zero-one loss function there is a faster inference approach by modifying the prediction algorithm
to compute additionally the second best output and taking the best result according to the value of objective function.\\

\noindent \textbf{Hamming Distance/Hamming Loss} ($\Delta_{\textit{HD}}, \Delta_{\textit{HL}}$)\\
\noindent In the context of sequence learning, given a true output $\bfy^*$ and a prediction $\bfy$ of the same length, \emph{Hamming distance} measures the number of states on which the two sequences disagree:
\begin{equation}
\label{L_HD}
\Delta_{\textit{HD}}(\bfy^*, \bfy) = \sum_{t = 1}^{M} \mathbb{1}_1[\bfy_t^* \neq \bfy_t].
\end{equation}
By normalising this value
we get \emph{Hamming loss} which does not depend on the length of the sequences. Both measures are decomposable.\\

\noindent \textbf{Weighted Hamming Distance} ($\Delta_{\textit{WHD}}$)\\
\noindent For a given matrix $\text{weight} \in \mathbb{R}^{N \times N}$, \emph{weighted Hamming distance} is defined according to $\Delta_{\textit{WHD}}(\bfy^*, \bfy) = \sum_{t=1}^M \text{weight}(y_t^*, y_t)$. Keeping track of the accumulated sum of the weights until the current position $t$ in a sequence (unlike for Hamming distance) can be intractable. We can use, however, the following observation. It is sufficient to count the numbers of occurrences $(\bfy_t^*,\bfy_t)$ for each pair of states $\bfy_t^*,\bfy_t \in \{1, ..., N\}$ according to
\begin{equation}
\begin{aligned}
& \sum_{t = 1}^{M} \text{weight}(y_t^*, y_t) =\\
& \sum_{s_1, s_2} \text{weight}(s_1,s_2) \sum_{t=1}^{M} \mathbb{1}_1[y_t^* = s_1 \land y_t = s_2].
\end{aligned}
\end{equation}
That is, each dimension of $\bfG$ (denoted $G_{s_1,s_2}$) corresponds to
\begin{equation}
G_{s_1,s_2}(\bfy; \bfy^*) = \sum_{t=1}^{M} \mathbb{1}_1[y_t^* = s_1 \land y_t = s_2].
\end{equation}
Here, we can upper bound the image size of $\bfG(\cdot)$ by considering an
urn problem with $N^2$ distinguishable urns and $M$ indistinguishable
balls. The number of possible distributions of balls over the urns is given by $\binom{M+N^2-1}{M} \leqslant M^{N^2}$.\\

\noindent \textbf{False Positives/Precision/Recall} ($\Delta_{\textit{\#FP}}, \Delta_{\textit{P}}, \Delta_{\textit{R}}$)\\
\noindent \emph{False positives} measure the discrepancy between outputs by counting
the number of false positives in a prediction $\bfy$ with respect to the true output $\bfy^*$ and has been often used
in learning tasks like natural language parsing due to its simplicity.
Precision and recall are popular measures used in information retrieval. By subtracting the corresponding values from one we can easily convert them to a loss function. Unlike for precision given by $\textit{TP}/(\textit{TP} + \textit{FP})$, recall effectively depends only on one parameter. Even though it is originally parameterised by two parameters given as $\textit{TP}/(\textit{TP} + \textit{FN})$ we can exploit the fact that the value $|\bfy^*| = \textit{TP} + \textit{FN}$ is always known in advance
during the inference rendering recall a decomposable measure.\\

\noindent \textbf{$F_{\beta}$-Loss} ($\Delta_{F_{\beta}}$)\\
\noindent \emph{$F_{\beta=1}$-score} is often used to evaluate the resulting performance in various natural language processing applications and is also appropriate for many structured prediction tasks. Originally, it is defined as the harmonic mean of precision and recall
\begin{equation}
F_{1} = \frac{2\textit{TP}}{2\textit{TP} + \textit{FP} + FN}.
\end{equation}
However, due to the fact that the value $|\bfy^*| = \textit{TP} + \textit{FN}$ is always known in advance during the inference, $F_{\beta}$-score effectively depends only on two parameters $(\textit{TP}, \textit{FP})$.
The corresponding loss function is defined as $\Delta_{F_{\beta}} = 1 - F_{\beta}$.\\

\noindent \textbf{Intersection Over Union} ($\Delta_{\cap/\cup}$)\\
\noindent \emph{Intersection Over Union} loss is mostly used
in image processing tasks like image segmentation
and object recognition and was used as performance measure
in the Pascal Visual Object Classes Challenge \cite{EveringhamEGWWZ15}.
It is defined as $1 - area(\bfy^* \cap \bfy)/area(\bfy^* \cup \bfy)$.
We can easily interpret this value in case where the outputs $\bfy^*$, $\bfy$ describe bounding boxes of pixels. The more the overlap of two boxes the smaller the loss value. In terms of contingency table this yields\footnote[1]{Note that in case of the binary image segmentation,
for example, we have a different interpretation of true and false positives. In particular, it holds $\textit{TP} + \textit{FN} = \textit{P}$, where $\textit{P}$ is the number of positive entries in $\bfy^*$.}
\begin{equation}
\Delta_{\cap/\cup} = 1 - \frac{\textit{TP}}{(\textit{TP}+\textit{FP}+\textit{FN})}.
\end{equation}
Since $|\bfy^*| = \textit{TP} + \textit{FN}$,
the value $\Delta_{{\cap}/{\cup}}$ effectively depends only on two parameters (instead of three). Moreover, unlike \textbf{$F_{\beta}$-loss}, $\Delta_{\cap/\cup}$ defines a proper distance metric on sets.\\

\noindent \textbf{Label-Count Loss} ($\Delta_{\textit{LC}}$)\\
\noindent \emph{Label-Count} loss is a performance measure which has been used for the task of binary image segmentation in computer vision and is given by
\begin{equation}
\Delta(\bfy^*, \bfy) = \frac{1}{M}\left|\sum_{i=1}^M y_i - \sum_{i=1}^M y_i^*\right|.
\end{equation}
This loss function prevents assigning low energy to segmentation labelings with
substantially different area compared to the ground truth.\\

\noindent \textbf{Crossing Brackets Number/Rate} ($\Delta_{\#CB}$, $\Delta_{\textit{CBR}}$)\\
\noindent The Number of \emph{Crossing Brackets} ($\#CB$) is a measure which has been used to evaluate the performance in natural language parsing by computing the average of how many constituents in one tree $\bfy$ cross over constituents boundaries in the other tree $\bfy^*$. The normalised version (by $|\bfy|$) of this measure is called \emph{Crossing Brackets (Recall) Rate}. Since the value $|\bfy|$ is not known in advance the evaluation requires a further parameter for the size of $\bfy$.\\

\noindent \textbf{Bilingual Evaluation Understudy} ($\Delta_{\textit{BLEU}}$)\\
\noindent \emph{Bilingual Evaluation Understudy} or for short \emph{BLEU} \cite{DBLP:conf/acl/PapineniRWZ02} is a measure which has been introduced to evaluate the quality of machine translations. It computes the geometric mean of the precision $p_k = TP_k/(TP_k+FP_k)$ of k-grams of various lengths (for $k = 1, ..., K$) between a hypothesis and a set of reference translations multiplied by a factor $BP(\cdot)$ to penalize short sentences according to
\begin{equation}
\Delta_{\textit{BLEU}}(\bfy^*, \bfy) = 1 - \textit{BP}(\bfy) \cdot \exp \left(\frac{1}{K}\sum_{k = 1}^K \log p_k\right).
\end{equation}
Note that $K$ is a constant rendering the term $M^{2K}$ a polynomial in $M$.\\

\noindent \textbf{Recall Oriented Understudy for Gisting Evaluation} \\($\Delta_{\textit{ROUGE-K}}$, $\Delta_{\textit{ROUGE-LCS}}$)\\
\noindent \emph{Recall Oriented Understudy for Gisting Evaluation} or for short \emph{ROUGE} \cite{DBLP:conf/grc/HeCMGLSW08} is a measure which has been introduced to evaluate the quality of a summary by comparing it to other summaries created by humans. More precisely, for a given set of reference summaries $Ref$ and a summary candidate $X$, ROUGE-K computes the percentage of k-grams from $Ref$ which appear in $X$
according to
\begin{equation}
\begin{aligned}
& \textit{ROUGE-K}(X, Ref) =\\
& \frac{\sum_{S \in Ref} \sum_{k \in \text{k-grams}(S)} \min\{\text{count}(k,X), \text{count}(k,S)\}}{\sum_{S \in Ref} \sum_{k \in \text{k-grams}(S)} \text{count}(k,S)}
\end{aligned}
\end{equation}
where $\text{count}(k,S)$ gives the number of occurrences of a k-gram $k$ in  a summary $S$.
We can estimate an upper bound $R$ on the image size of $\bfG(\cdot)$ similarly to the derivation for the weighted Hamming distance above as $M^D$ where $D := |\text{grams(Ref)}|$ is the dimensionality of $\bfG(\cdot)$, that is, the number of unique k-grams occurring in the reference summaries. Note that we do not need to count grams which do not occur in the references.

Another version \textit{ROUGE-LCS} is based on the concept of the longest common subsequence (LCS). More precisely, for two summaries $X$ and $Y$, we first compute $LCS(X,Y)$, the length of the LCS, which we then use to define some sort of precision and recall given by $LCS(X,Y)/|X|$ and $LCS(X,Y)/|Y|$, respectively. The latter two are used to evaluate a corresponding $F$-measure:
\begin{equation}
\begin{aligned}
& \Delta_{\textit{ROUGE-LCS}} =\\
& 1 - \frac{1}{|Ref|}\sum_{S \in Ref}\frac{(1+\beta^2)P_{LCS(X,S)}\cdot R_{LCS(X,S)}}{\beta^2P_{LCS(X,S)}+R_{LCS(X,S)}}.
\end{aligned}
\end{equation}
That is, each dimension in $\bfG(\cdot)$ is indexed by an $S \in Ref$.
\textit{ROUGE-LCS} (unlike \textit{ROUGE-K}) is non-decomposable.\\
\section{Related Works}
\label{sec6}
%When we drop the requirement on the worst-case running time to be polynomially bounded, there is a plethora of exact algorithms which are still fast in practice \cite{HurleyOAKSZG16,HallerSS18,SavchynskyyKSS13,KappesSRS13}. Our focus in this paper, however, is on the exact MAP inference which on one hand has polynomial run-time guarantee, and on the other hand allows for models with global dependencies which can be expressed by means of global cardinality-like potentials. Note that the latter renders the treewidth of the resulting problem unbounded.

Several previous works address  exact MAP inference with global factors in the context of  SSVMs when optimising for non-decomposable loss functions.
Joachims \cite{Joachims05asupport} proposed an algorithm
for a set of multivariate losses including $F_{\beta}$-loss.
However, the presented idea applies only to a simple case where the corresponding mapping $F$ in (\ref{Objective2})
decomposes into \emph{non-overlapping} components (e.\ g.\ unary potentials).

Similar ideas based on introduction of auxiliary variables
have been proposed \cite{tarlow2012fast, mezuman2013tighter}
to modify a belief propagation algorithm according to a special form of high order potentials.
Precisely, in case of binary-valued variables $y_i \in \{0,1\}$, the authors focus on the \emph{univariate} \emph{cardinality potentials}
$\eta(\bfG(\bfy)) = \eta(\sum_{i} y_i)$.
For the tasks of sequence tagging and constituency parsing \cite{BauerGBMK14, BauerBM17}
proposed an exact inference algorithm for the slack scaling formulation
focusing on \emph{univariate}
dissimilarity measures $\bfG(\bfy) = \sum_{t} \mathbb{1}_1[y_t^* \neq y_t]$ and $\bfG(\bfy) = \#\textit{FP}(\bfy)$ (see Table \ref{tab:DissilarityMeasure} for details).
In \cite{BauerSM17} the authors extrapolate this idea and provide a unified strategy to tackle multivariate and non-decomposable loss functions.

In the current paper we build upon the results in \cite{BauerSM17} and generalise the target problem (Problem \ref{mainproblem})
increasing the range of admissible applications. 
More precisely, we replace the binary operation $\odot \colon \mathbb{R} \times \mathbb{R} \rightarrow \mathbb{R}$ corresponding to either a summation or  multiplication in the previous objective $F(\bfy) \odot \eta(\bfG(\bfy))$ by a function $H \colon \mathbb{R} \times \mathbb{R}^P \rightarrow \mathbb{R}$ allowing for more subtle interactions between the energy of the core model $F$ and the sufficient statistics $\bfG$ according to $H(F(\bfy), \bfG(\bfy))$. The increased flexibility, however, must be further restricted in order for a corresponding solution to be optimal.  We found that it is sufficient to impose a requirement on $H$ to be non-decreasing in the first argument.
Note that in the previous objective \cite{BauerSM17} this requirement automatically holds.

%Furthermore, previous work only guarantees polynomial run-time in the case where the core model can be represented by a tree-shaped factor graph excluding problems with cyclic dependencies. Here, we complete the analysis by extending this result to the more general case of clique trees. We also improve upon the estimation of the upper bound on the computational complexity from $O(M \cdot N^{\tau+1} \cdot {R}^{{\nu}-1})$ with potentially unbounded $\nu$ to $O(M \cdot N^{\tau+1} \cdot {R}^{2})$.
Furthermore, previous work only guarantees polynomial run-time in the case where a) the core model can be represented by a tree-shaped factor graph and b) the maximal degree of a variable node in the factor graph is bounded by a constant. The former excludes problems with cyclic dependencies and the second rejects graphs with star-like shapes. A corresponding idea  for clique trees can handle cycles but suffers from a similar restriction on the maximal node degree to be bounded.
Here, we solve this problem by applying a graph transformation proposed in Section \ref{sec:LMP} effectively reducing the maximal node degree to $\nu = 3$. We note that a similar idea can be applied to factor graphs by replicating variable nodes and introducing constant factors. Finally, we improve upon the guarantee on the computational complexity by reducing potentially unbounded  parameter $\nu$ in the upper bound $O(M \cdot N^{\tau+1} \cdot {R}^{{\nu}-1})$ to $\nu \leqslant 3$.
\section{Validation of Theoretical Time Complexity}
\label{sec7}
\begin{figure}[t]
\centering
\includegraphics[scale = 0.69]{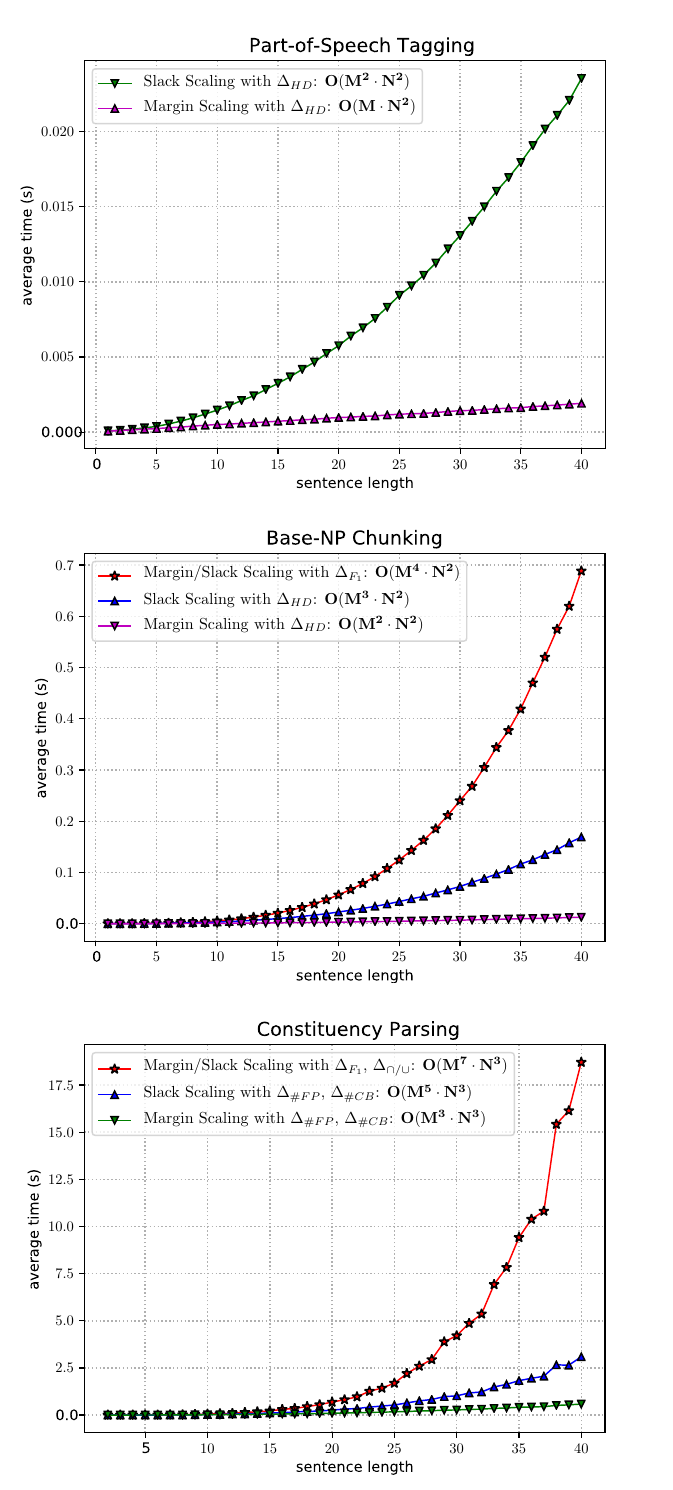}
\caption{Empirical evaluation of the run-time performance for the task of loss augmented inference on example of applications in part-of-speech tagging, base-NP chunking and constituency parsing.
}
\label{fig_plots}
\end{figure}

To demonstrate feasibility, we evaluate the performance of our algorithm on several application tasks: part-of-speech tagging \cite{DBLP:conf/cicling/Manning11}, base-NP chunking \cite{DBLP:conf/coling/TongchimSI08} and constituency parsing \cite{Tsochantaridis05largemargin, BauerBM17}.
More precisely, we consider the task of loss augmented inference (Section \ref{sec4_a}) for margin and slack scaling with different loss functions. The run-times for the tasks of diverse K-best MAP inference (Section \ref{sec4_c}) and for evaluating structured generalisation bounds (Section \ref{sec4_b}) are identical with the run-time for the loss augmented inference with slack scaling. We omit the corresponding plots due to redundancy.
In all experiments we used  the Penn English Treebank-3 \cite{MarcusKMMBFKS94} as a benchmark data set which provides a large corpus of annotated sentences from the Wall Street Journal.
For better visualisation we restrict our experiments to sentences containing at most 40 words. The resulting time performance is shown in Figure \ref{fig_plots}.
We can see that different loss functions result in different computation costs depending on the number of values
for auxiliary variables given in the last column of Table \ref{tab:DissilarityMeasure}.
In particular, the shapes of the curves are consistent with the upper bound provided in Theorem \ref{theorem2} reflecting the polynomial degree of the overall dependency with respect to the graph size $M$.
The difference between margin and slack scaling originates from the fact that in the case of a decomposable loss function $\bfG$ the corresponding loss terms can be folded into the factors of the compatibility function $F$ allowing for the use of conventional message passing.

\section{Conclusion}
\label{sec8}
Despite the high diversity in the range of existing applications, a considerable number of the underlying MAP problems share the same unifying property that the information on the global variable interactions imposed by either a global factor or a global constraint can be locally  propagated trough the network by means of the dynamic programming. 
Extending previous work we presented a theoretical framework for efficient exact inference in such a case and showed that the tractability assumption of junction tree algorithm on the model for a corresponding treewidth to be bounded is only a sufficient condition for an efficient inference which can be further relaxed depending on the form of global connections. In particular, the performance of our approach does not explicitly depend
on the graph structure but rather on intrinsic properties like treewidth and number of states of the auxiliary variables defined by the sufficient statistics of global interactions.
The overall computational procedure is provable exact and has lower asymptotic bounds on the computational time complexity compared to the previous work.
\appendices
\section{Proof of Theorem \ref{theorem2}}
\label{sec:Proof2}
\begin{proof}
We now show the correctness of the presented computations.
For this purpose we first provide a semantic interpretation of messages as follows.
Let $C_i - C_j$ be an edge in a clique tree.
We denote by $\mcF_{\prec(i-j)}$ the set of clique factors $f_{C_k}$ of the mapping $F$ on the $C_i$-th side of the tree and by $\mcG_{\prec(i-j)}$ a corresponding set of clique factors of the mapping $\bfG$.
Furthermore, we denote by $\mcV_{\prec(i-j)}$ the set of all variables appearing on the $C_i$-th side but not in the sepset $C_i \cap C_j$.
Intuitively, a message $\mu_{C_i \rightarrow C_j}^{\bfl_i}(\bfy_{C_i \cap C_j})$ sent from a clique $C_i$ to $C_j$ corresponds to a sum of all factors
contained in $\mcF_{\prec(i-j)}$ which is maximised (for fixed values of $\bfy_{C_i \cap C_j}$ and ${\bfl_i}$) over the variables in $\mcV_{\prec(i-j)}$
subject to the constraint $\bfl_i = \sum_{\bfg_{C_k} \in \mcG_{\prec(i-j)}} \bfg_{C_k}(\bfy_{C_k})$.
That is, we define the following induction hypothesis
\begin{equation}\label{claim1}
\begin{aligned}
& \mu_{C_i \rightarrow C_j}^{\bfl_i}(\bfy_{C_i \cap C_j}) =\\
& \underset{\mcV_{\prec(i-j)}\colon \bfl_i = \sum_{\bfg_{C_k} \in \mcG_{\prec(i-j)}} \bfg_{C_k}(\bfy_{C_k})}{\max} \sum_{f_{C_k} \in \mcF_{\prec(i-j)}} f_{C_k}(\bfy_{C_k}).
\end{aligned}
\end{equation}
Now consider an edge $(C_i - C_j)$ such that $C_i$ is not a leaf.
Let $i_1, ..., i_m$ be the neighbouring cliques of $C_i$ other than $C_j$.
It follows from the running intersection property
that $\mcV_{\prec(i-j)}$ is a disjoint union of $\mcV_{\prec(i_k-i)}$ for $k = 1, ..., m$
and the variables $\bfy_{C_i \setminus C_j}$ eliminated at $C_i$ itself.
Similarly, $\mcF_{\prec(i-j)}$ is the disjoint union of the $\mcF_{\prec(i_k-i)}$
and $\{f_{C_i}\}$.
Finally, $\mcG_{\prec(i-j)}$ is the disjoint union of the $\mcG_{\prec(i_k-i)}$
and $\{\bfg_{C_i}\}$.
In the following,
we abbreviate the term $\mcV_{\prec(i_k-i)}\colon \bfl_{i_k} = \sum_{\bfg \in \mcG_{\prec(i_k-i)}} \bfg$ describing a range of variables in $\mcV_{\prec(i_k-i)}$ subject to a corresponding equality constraint with respect to $\bfl_{i_k}$ by $\mcV_{\prec(i_k-i)}\colon \bfl_{i_k}$.
Thus, the right hand side of Eq. (\ref{claim1}) is equal to
\begin{equation}\label{E1}
\begin{aligned}
\small
\max_{\bfy_{C_i \setminus C_j}} \max_{\{\bfl_{i_k}\}_{k=1}^m} \max_{\mcV_{\prec(i_1-i)}\colon \bfl_{i_1}} \cdots \max_{\mcV_{\prec(i_m-i)}\colon \bfl_{i_m}} \left( \sum_{f \in \mcF_{\prec(i_1-i)}} f \right)\\
+ \cdots + \left( \sum_{f \in \mcF_{\prec(i_m-i)}} f \right) + f_{C_i}
\end{aligned}
\end{equation}
where in the second $\max$ we maximise over all configurations of $\{\bfl_{i_k}\}_{k=1}^m$
subject to the constraint $\sum_{k=1}^m \bfl_{i_k} = \bfl_i - \bfg_{C_i}(\bfy_{C_i})$.
Since all the corresponding sets are disjoint the term (\ref{E1}) is equal to
\begin{equation}\label{E2}
\begin{aligned}
\small
\max_{\bfy_{C_i \setminus C_j}, \{\bfl_{i_k}\}_{k=1}^m} f_{C_i} + \underbrace{\max_{\mcV_{\prec(i_1-i)}\colon \bfl_{i_1}} \left( \sum_{f \in \mcF_{\prec(i_1-i)}} f \right)}_{\mu_{C_{i_1} \rightarrow C_i}^{\bfl_{i_1}}(\bfy_{C_{i_1} \cap C_i})}+\\
\cdots + \underbrace{\max_{\mcV_{\prec(i_m-i)}\colon \bfl_{i_m}} \left( \sum_{f \in \mcF_{\prec(i_m-i)}} f \right)}_{\mu_{C_{i_m} \rightarrow C_i}^{\bfl_{i_m}}(\bfy_{C_{i_m} \cap C_i})}
\end{aligned}
\end{equation}
where again the maximisation over $\{\bfl_{i_k}\}_{k=1}^m$
is subject to the constraint $\sum_{k=1}^m \bfl_{i_k} = \bfl_i - \bfg_{C_i}(y_{C_i})$.
Using the induction hypothesis in the last expression we get the right hand side of Eq. (\ref{E_mp_})
proving the claim in Eq. (\ref{claim1}).

Now look at Eq. (\ref{beliefs_}).
Using Eq. (\ref{claim1}) and the fact that all involved sets of variables and factors for different messages are disjoint
we conclude that the computed values $\mu(\bfl)$ corresponds to the sum of all factors $f$ for the mapping $F$ over the variables in $\bfy$ which is maximised subject  to the constraint $\bfG(\bfy) = \bfl$.
Note that until now the proof is equivalent to the one given for Theorem 2 in the previous publication \cite{BauerSM17} because the message passing part
constrained via auxiliary variables is identical. Now we use an additional requirement on $H$ to ensure optimality of a corresponding solution.
Because $H$ is non-decreasing in the first argument, by performing maximisation over all values $\bfl$ according to Eq. (\ref{solution_}) we get the optimal value of Problem \ref{mainproblem}.

By inspecting the formula for the message passing in Eq. (\ref{E_mp_}) we conclude that the corresponding operations
can be done in $O(M \cdot N^{\tau+1} \cdot R^{\nu-1})$ time
where $\nu$ denotes the maximal number of neighbors	 of any clique node $C_i$.
First, the summation in Eq. (\ref{E_mp_}) involves $|ne(C_i)|$ terms resulting in $|ne(C_i)|-1$ summation operations.
Second, a maximisation is performed first over $|C_i\setminus C_j|$ variables with a cost $N^{|C_i\setminus C_j|}$.
This, however, is done for each configuration of $\bfy_{C_i \cap C_j}$
where $|C_i \setminus C_j| + |C_i \cap C_j| = |C_i| \leqslant \tau + 1$ resulting in $N^{\tau+1}$.
Then a maximisation over $\{\bfl_k\}$ costs additionally $R^{|ne(C_i)|-2}$. Together with possible values for $\bfl$
it yields $R^{\nu-1}$ where we upper bounded $|ne(C_i)|$ by $\nu$.
Therefore, sending a message for all possible configurations of $(\bfy_{C_i \cap C_j};{\bfl})$
on the edge $C_i - C_j$
costs $O(N^{\tau+1} \cdot (|ne(C_i)|-1) \cdot R^{\nu-1})$ time.
Finally, we need to do these operations for each edge $(i,j) \in E$ in the clique tree.
The resulting cost can be estimated as follows:
 $\sum_{(i,j) \in E} N^{\tau+1} \cdot R^{\nu-1} \cdot (|ne(C_i)|-1) =
N^{\tau+1} \cdot R^{\nu-1} \sum_{(i,j) \in E} (|ne(C_i)|-1) \leqslant N^{\tau+1} \cdot R^{\nu-1} \cdot |E|
= N^{\tau+1} \cdot R^{\nu-1} \cdot (|V|-1) \leqslant N^{\tau+1} \cdot R^{\nu-1} \cdot M$,
where $V$ denotes the set of cliques nodes in the clique tree.
%The time for constructing a valid clique tree is dominated by the time for message passing.
Therefore, the total complexity is upper bounded by $O(M \cdot N^{\tau+1} \cdot R^{\nu-1})$.
\end{proof}

\section{Proof of Proposition \ref{p_nu}}
\label{sec:Proof_prop}
\begin{figure}[t]
\centering
\includegraphics[scale = 0.7]{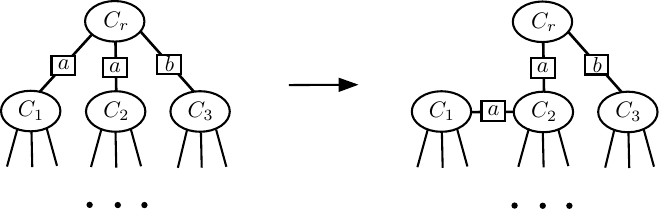}
\caption{Illustration of the reshaping procedure for a clique tree in the case where the condition $\nu \leqslant 2^{\tau+2}-4$ is violated. $C_r$ is the root clique where a sepset $a$ occurs at least two times.
The number of neighbors of $C_r$ can be reduced by removing the edge between $C_1$ and $C_r$ and attaching $C_1$ to $C_2$. This way we can ensure that every node has at most one duplicate for every possible sepset. Furthermore, this procedure preserves the running intersection property.
}
\label{fig_proposition_nu}
\end{figure}

\begin{proof}
Assume we are given a clique tree with treewidth $\tau$.
That is, every node in a clique tree has at most $\tau+1$ variables. Therefore, the number of all possible variable combinations for a sepset is given by $2^{\tau+1}-2$ where we exclude the empty set and the set containing all the variables in the corresponding clique.

Furthermore, we can deal with duplicates by rearranging the edges in the clique tree such that that for every sepset from the $2^{\tau+1}-2$ possibilities there is at most one duplicate resulting in $2^{\tau+2}-4$ possible sepsets. More precisely,
we first choose a node having more than $2^{\tau+2}-4$ neighbors as root and then reshape the clique tree by propagating some of the neighbors towards the leaves as illustrated in Figure \ref{fig_proposition_nu}. Due to these procedure the maximal number of repetition for every  sepset is upper bounded by $2$.
\end{proof}

\section{Proof of Theorem \ref{theorem3}}
\label{sec:Proof4}
\begin{proof}
We provide a proof per induction.
Let $C_1, ..., C_K$ be the cliques of a corresponding instance of Problem \ref{mainproblem}.
We now consider an arbitrary but fixed order of $C_k$ for $k \in \{1, ..., K\}$.
We denote by $R_i$ the number os states of a variable $\bfl_i$, that is, $R$ is given by $\max_{i} R_i$.
As previously mentioned (see equation (\ref{eq_aux})), an auxiliary variable $\bfl_i$ corresponds to a sum of potentials $\bfg_{C_k}(\bfy_{C_k})$
over all $C_k$ in a subtree of which $C_i$ is the root.
That is, the number $R_i$ is upper bounded by the image size of a corresponding sum-function according to
\begin{equation}
R_i \leqslant \underbrace{\left|\sum_{k = 1}^{i} \bfg_{C_k}(\cdot)\right|}_{\in [-i\cdot T, i\cdot T] \cap \mathbb{Z}} \leqslant 2 \cdot i \cdot T
\end{equation}
where the corresponding values are in the set $[-i\cdot T, i\cdot T] \cap \mathbb{Z}$ defining our induction hypothesis.
Using induction hypothesis and the assumption $\bfg_{C_k} \in [-T, T]$ it directly follows that the values for $R_{i+1}$ are all in the set $[-(i+1)\cdot T, (i+1)\cdot T] \cap \mathbb{Z}$.
The base case for $R_1$ holds due to the assumption of the theorem.
Because $K \leqslant M$ always holds and the order of the considered cliques is arbitrary, we can conclude that $R$ is upper bounded by $2\cdot M \cdot T$ which is a polynomial in $M$.
\end{proof}
%\noindent $\Box$


\ifCLASSOPTIONcaptionsoff
  \newpage
\fi



%\bibliographystyle{IEEEtran}

%\bibliographystyle{plain}
\bibliographystyle{IEEEtran}
\nocite{*}
\bibliography{references}

\end{document}